\documentclass[ reprint, 
 aps,
 pra,
 showkeys,
 nofootinbib
]{revtex4-1}
\usepackage{amsfonts}
\usepackage{amssymb}
\usepackage{amsmath}
\usepackage{graphicx}
\usepackage{subfigure}
\usepackage{dcolumn}
\usepackage{bm}
\usepackage{setspace}
\usepackage[hang]{footmisc}
\usepackage[sort&compress]{natbib}
\usepackage{multirow}
\usepackage[sort&compress]{cleveref}
\usepackage{nomencl}
\usepackage[english]{babel}
\usepackage{xcolor, soul}
\usepackage{marginnote}
\usepackage{comment}
 
\sethlcolor{yellow}

\crefname{equation}{\unskip}{\unskip}
\crefname{figure}{\unskip}{\unskip}
\crefname{section}{\unskip}{\unskip}
\crefname{subsection}{\unskip}{\unskip}
\begin{document}
\let\ref\cref

\title{Fluid leakage in metallic seals}

\author{F.J. Fischer}
\affiliation{IFAS, Aachen University, Germany}
\author{K. Schmitz}
\affiliation{IFAS, Aachen University, Germany}
\author{A. Tiwari}
\affiliation{PGI-1, FZ J\"ulich, Germany}
\affiliation{www.MultiscaleConsulting.com, Wolfshovener str. 2, 52428 J\"ulich}
\author{B.N.J. Persson}
\affiliation{PGI-1, FZ J\"ulich, Germany}
\affiliation{www.MultiscaleConsulting.com, Wolfshovener str. 2, 52428 J\"ulich}

\begin{abstract}
{\bf Abstract:} Metallic seals are crucial machine elements in many important applications, e.g.,
in ultrahigh vacuum systems. Due to the high elastic modulus of metals, and the 
surface roughness which exists on all solid surfaces, if no plastic deformation would occur
one expects in most cases large fluid flow channels 
between the contacting metallic bodies, and large
fluid leakage. However, in most applications plastic deformation occurs, at least at the asperity level,
which allows the surfaces to approach each other to such an extent that fluid leakage often can
be neglected. In this study we present an experimental set-up for studying the fluid leakage in
metallic seals. We study the water leakage between a steel sphere
and a steel body (seat) with a conical surface. 
The experimental results are found to be in good
quantitative agreement with a (fitting-parameter-free) theoretical model.  
The theory predicts that the plastic 
deformations reduce the leak-rate by a factor $\approx 8$.
\end{abstract}

\maketitle
\makenomenclature


\section{Introduction}

Seals are a crucial machine element used to confine 
a high pressure fluid to some given volume. Due to the
interfacial surface roughness
most seals exhibit leakage \cite{add1,armand}. To minimize the leakage, seals are usually made 
from a soft material, such as rubber (with an elastic modulus of order $E\approx 10 \ {\rm MPa}$), which
can easily deform elastically and reduce the gap to the 
counter surface to such an extent that the 
fluid leakage becomes negligible or unimportant.

For some applications, e.g., involving high temperatures or 
hot reactive gases, or very high fluid pressures, rubber-like
materials cannot be used. In these cases, and in ultra high vacuum systems, seals made from metals are
very useful \cite{thesis,thesis1,metals}. 

Metals are elastically very stiff 
(typical elastic modulus of order $E \approx 100 \ {\rm GPa}$), 
and unless the surfaces are extremely smooth, or the nominal contact pressure
extremely high, calculations (assuming purely elastic deformations) 
show that large non-contact channels would occur at the interface 
resulting in a large fluid leakage. However, most metals yield plastically
at relative low contact pressures, typically of order $\sigma_{\rm Y} \approx 1 \ {\rm GPa}$.
This will allow the contacting surfaces to approach each 
other, which will reduce
the interfacial gap to such an extent that the fluid leakage usually can be neglected.

For purely elastic solids like rubber, contact mechanics theories have been developed for how to predict the fluid 
leak-rate, and it has been shown that they are in good agreement with experiments \cite{Lorenz1,Lorenz2}. 
The simplest approach assumes that the whole fluid pressure difference between the inside and outside of the sealed region occurs over 
the most narrow constrictions (denoted critical junctions) which are encountered along the largest open percolating non-contact flow channels.

For elastic solids numerical contact mechanics models \cite{Ref7}, such as the boundary element model, 
and the analytic theory of Persson \cite{BP,Alm}, can be used to calculate the surface separation at the critical 
junction and hence predict fluid leakage rates. For solids exhibiting plastic flow, the surfaces will approach each other 
more closely than if only elastic deformations would occur. This will reduce the fluid leakage rate \cite{Per1,AA4}.

We have recently shown how the leakage of static rubber seals can be estimated using the
Persson contact mechanics theory combined with the Bruggeman effective medium theory \cite{Yang,Review,Boris,LP1,liftoff} 
(for other approaches, see Ref. \cite{Carbone,Mueser,Mus1,Mus2,metals}). In this paper
we apply the theory to metallic seals where plastic deformations are important unless the surfaces
are extremely smooth \cite{Pei,Kadin,Zhao,tobe}.

Experimental studies of plastic deformation of rough metallic and polymeric surfaces was presented in Ref. \cite{Av,Av2}.
Several studies of surface roughness and plastic flow have been reported using microscopic (atomistic) models \cite{Pas}, 
or models inspired by atomic scale phenomena that
control the nucleation and glide of the dislocations \cite{Nic1,Nic2,Nic3,Nic4}. 
These models supply fundamental insight into the complex process of plastic flow, but are not easy to 
apply to practical systems involving inhomogeneous polycrystalline metals and alloys exhibiting surface roughness 
of many length scales. 
The approach used in this study is less accurate but easy to implement, and it can be used to 
estimate the leakage rates of metallic seals. 

\section{Experimental}

\subsection{Surface topography}

The topography measurements were performed with a Mitutoyo Portable Surface Roughness Measurement device, 
Surftest SJ-410 with a diamond tip with the radius of curvature $R = 1 \ {\rm \mu m}$, and with 
the tip--substrate repulsive force $F_{\rm N}=0.75 \ {\rm mN}$. The lateral tip speed was $v=50 \ {\rm \mu m/s}$.

From the measured surface topography (line scans) $z=h(x)$ we calculated the one-dimensional (1D) surface roughness power spectra defined by
$$C_{\rm 1D} (q) = {1\over 2 \pi} \int_{-\infty}^\infty dx \ \langle h(x) h(0) \rangle e^{i q x}$$
where $\langle .. \rangle$ stands for ensemble averaging.
For surfaces with isotropic roughness, the 2D power spectrum $C(q)$ can be obtained directly from $C_{\rm 1D} (q)$
as described elsewhere \cite{Review,fractal,Nyak,CarbLor}. For randomly rough surfaces, all the (ensemble averaged) information about 
the surface is contained in the power spectrum $C(q)$. For this reason the only information about the surface roughness 
which enter in contact mechanics theories (with or without adhesion) is the function $C(q)$. 
Thus, the (ensemble averaged) area of real contact, the interfacial stress distribution and the distribution of 
interfacial separations, are all determined by $C(q)$ \cite{BP,Alm,C5,Yang1,PRL}.

\begin{figure}
\includegraphics[width=0.5\columnwidth]{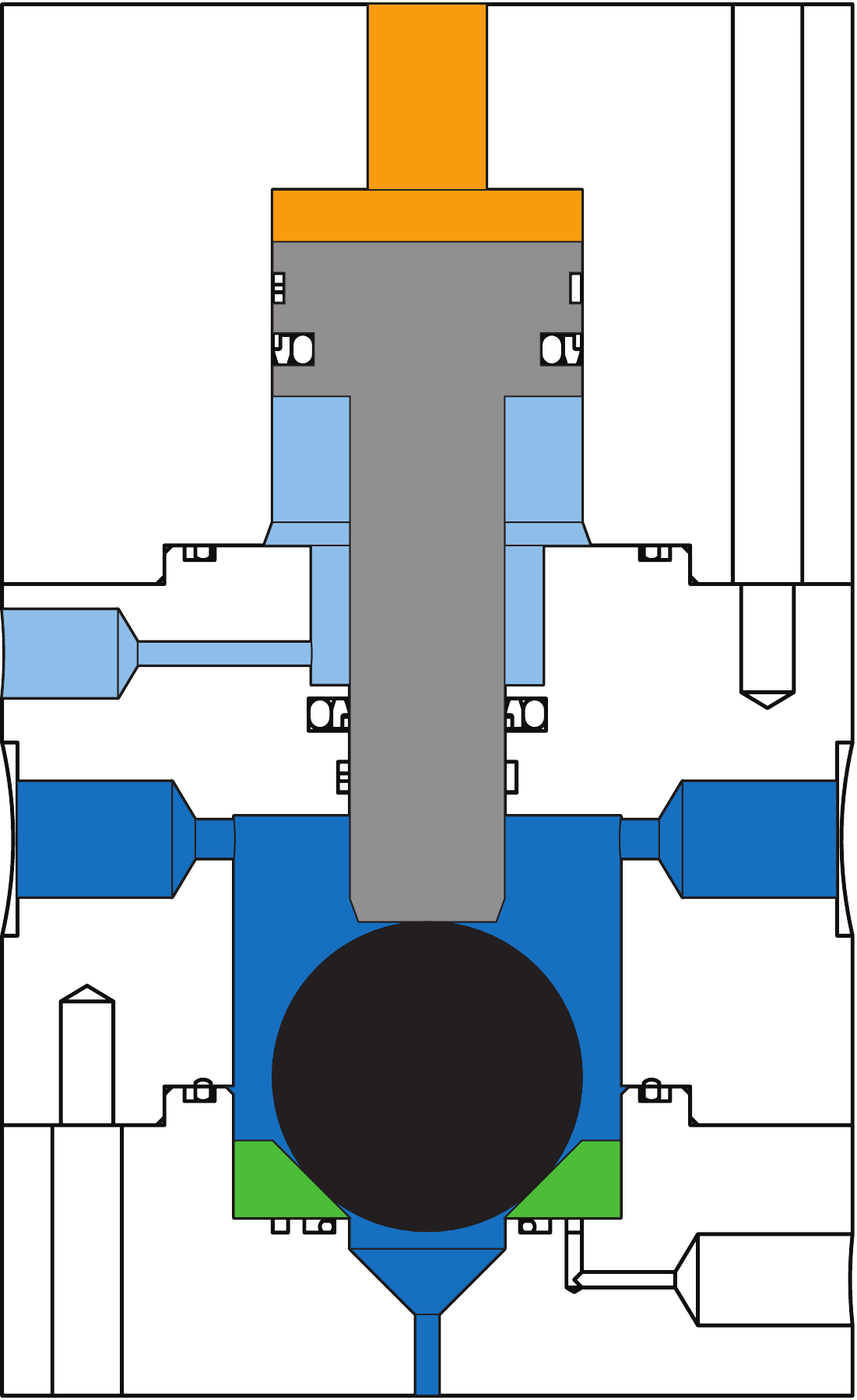}
\caption{\label{ballseatAachen.pdf}
Schematic picture of the leakage experiment. The dark blue color is
water, the light blue color is air and the yellow color is oil. The steel ball (black)
is squeezed against the steel seat (green) in part by the applied water pressure, and in part by the 
steel piston contacting it from on-top (gray cylinder) using an oil based hydraulics system.
In the present experiments no force has been applied by the steel piston.
}
\end{figure}

    \subsection{Leak-rate experiment}

    The aim of the experiment is to measure the fluid (here water) leakage for a ball-seat 
    valve as a function of the applied fluid pressure.
    In the present set-up we can apply fluid pressures up to $20 \ {\rm bar}$,
    and the pressure can be kept at a constant level even if there is leakage.
    In this study we are interested in the influence of the surface roughness on the leak-rate.
    The steel ball we use is very smooth but we use different seats with varying surface roughness
    produced by sandblasting. Thus the seat must be easily replaceable.

    In the experiments reported on below, the ball is squeezed against the seat only by the fluid pressure.
    However, the experimental set-up also includes the ability to create an additional normal load onto the ball. 
    This additional force is generated by pushing a steel piston against the ball using an oil based hydraulics system.

    The test chamber, which contains the ball and the seat, can be seen in Fig.~\ref{ballseatAachen.pdf}.
    Here, the chamber surrounding the ball is filled with water (dark blue color).
    Purified water is used as the leakage fluid because the low viscosity increases the 
    leakage which makes the measurement more easy and accurate as compared to using a hydraulic oil.
    Purified water has very few contamination particles which can clog the flow channels.
    However, if water is used in a hydraulics system one has to be careful to avoid corrosion 
    of all surfaces, including the seat and the ball.

    The method used to determine the leak-rate depends on the amount of leakage.
    For very low amounts of leakage the best way is to count the amount of water drops over time.
    The volume of a single drop can be estimated by repeated measurement of multiple drops.
    For higher amounts of leakage a measuring cylinder can be used.
    If the leakage surpasses the typical volume of a measuring cylinder, the leakage is instead determined by measuring its mass using a scale.

\begin{figure}
\includegraphics[width=0.5\columnwidth]{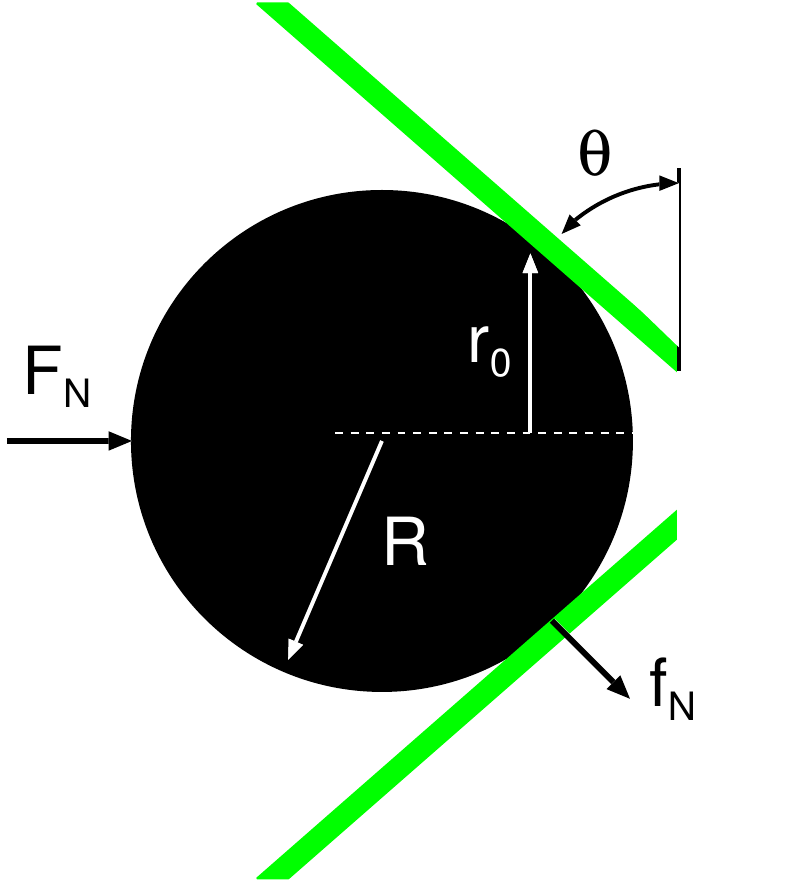}
\caption{\label{ballSEAL.pdf}
A steel ball (radius $R$) squeezed against a conical surface.
The radius of the ball $R=2 \ {\rm cm}$ and the cone angle $\theta = 45^\circ $.
In the experiments, the axial force $F_{\rm N}$ squeezing the ball against the cone surface
is due only to the fluid pressure difference between inside and outside the seal 
so that $F_{\rm N} = \pi r_0^2 p_{\rm fluid}$.
}
\end{figure}


\begin{figure}
\includegraphics[width=1.0\columnwidth]{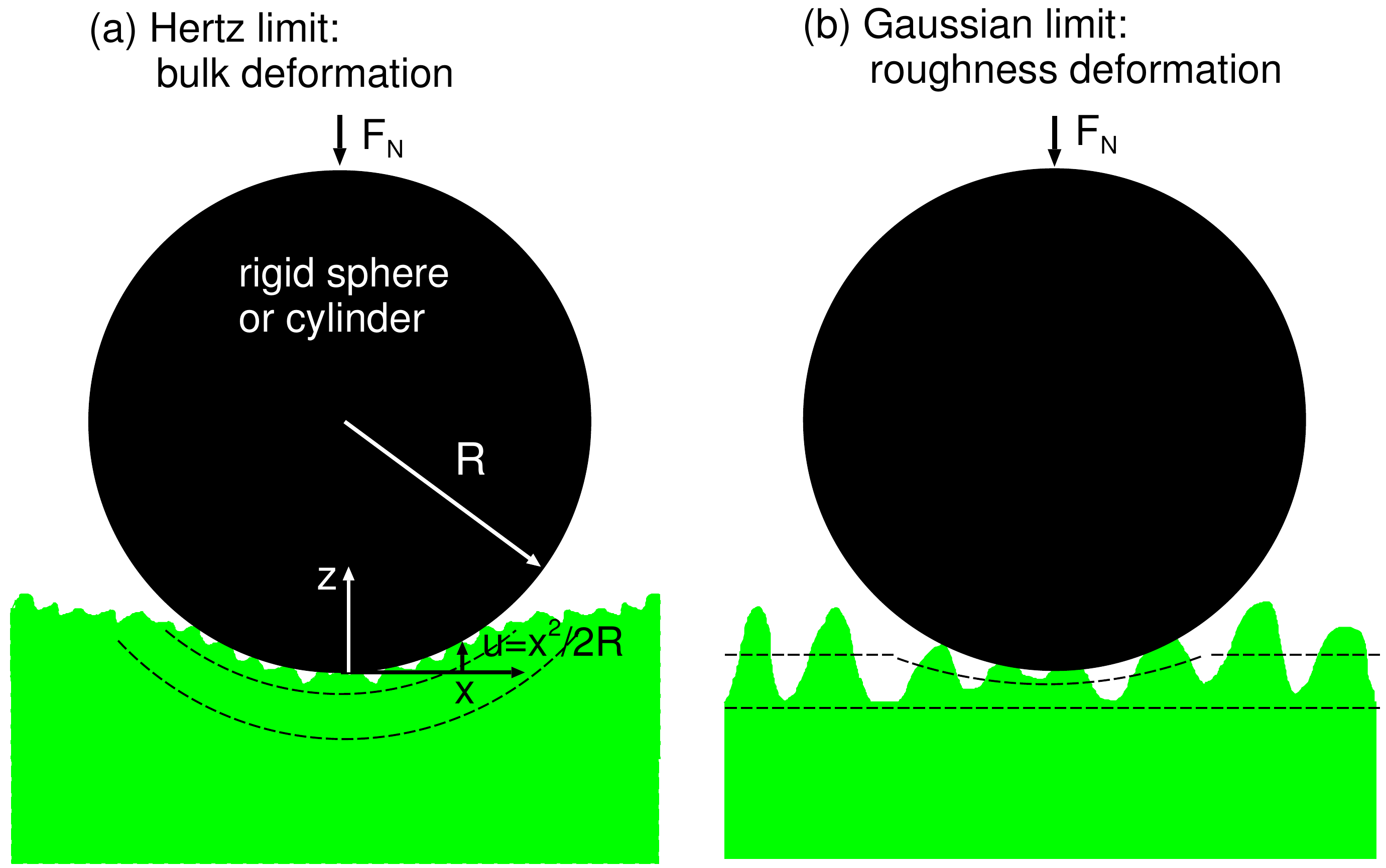}
\caption{\label{TwoLimits.pdf}
Two limiting cases when a rigid cylinder (black) with radius $R$ is squeezed against a nominal flat halfspace (green). (a) If the surface roughness amplitude is very small, or the applied force very high, the nominal contact area will be determined by bulk deformations and given by the Hertz contact theory. (b) In the opposite limit mainly the surface asperities deform (but with a long-range elastic coupling occuring between them). In this limit the pressure profile is Gaussian- like.
}
\end{figure}


\begin{figure}\includegraphics[width=0.95\columnwidth]{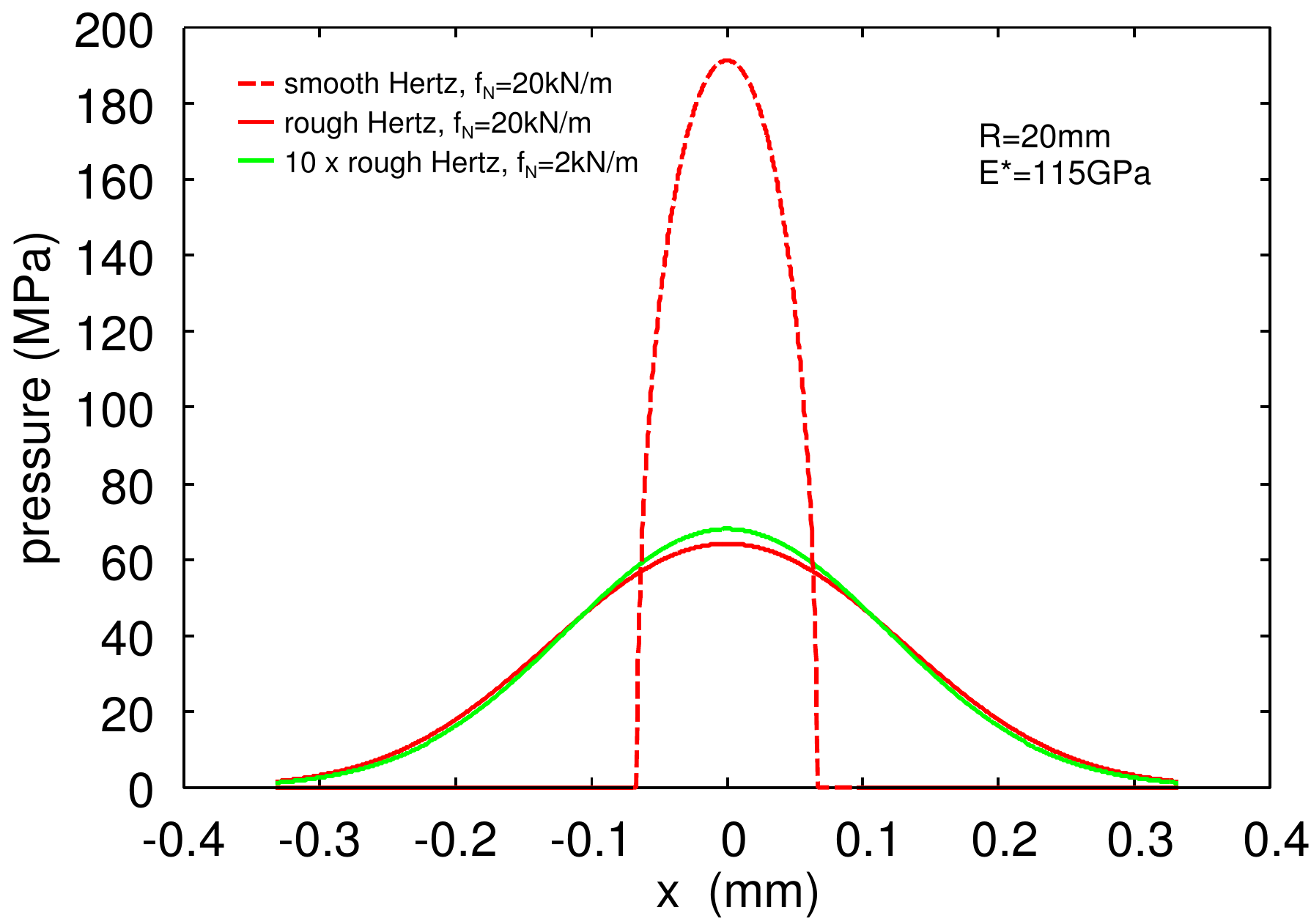}
\caption{\label{1x.2pcont.used.region.pdf}
The calculated pressure distributions 
for a smooth seat surface (red dashed line), and for the rough seat surface (red solid line) for the 
line load $f_{\rm N}=20 \ {\rm kN/m}$ (corresponding to the fluid pressure $p_{\rm fluid}=20 \ {\rm bar}$). 
For the rough surface we also show the pressure profile for the
line load $f_{\rm N}=2 \ {\rm kN/m}$ (corresponding to the fluid pressure $p_{\rm fluid}=2 \ {\rm bar}$),
multiplied by a factor of $10$. In the calculations we used
and the effective elastic modulus $E^*=115 \ {\rm GPa}$. 
}
\end{figure}

\section{Leakage Calculations}

The calculation of the fluid leakage in metallic seals involves several steps.
First it is necessary to determine the nominal contact pressure profile $p(x,y)$ acting 
at the interface between the two metallic bodies. This will in general
involve both elastic and plastic deformations of the metals. Secondly, one must determine
the separation $u(x,y)$ between the surfaces as this will determine the fluid flow channels
at the interface. This problem will depend on the pressure profile $p(x,y)$
and on the surface roughness, and the elastoplastic properties of the metals.
Finally, one must calculate the fluid flow at the interface in the open (non-contact) channels
from the high pressure side to the low pressure side. This is a complex hydrodynamic
problem which in general cannot be solved exactly. 

\subsection{Contact Force}

The experimental set-up consists of a steel ball (radius $R$) 
and a conical steel body (seat) with the angle $\theta$ defined in Fig. \ref{ballSEAL.pdf}.
A fluid with the pressure $p_{\rm fluid}$ squeezes the ball against the seat. 
We assume here that there is no other applied force squeezing the ball against the seat.
The contact region between the ball and the seat forms a circular region (line contact) with radius $r_0$
(see Fig.~\ref{ballSEAL.pdf}). Hence, the force squeezing the ball against the seat is
    $F_{\rm N} = \pi r_0^2 p_{\rm fluid}$. The force per unit circumferential length is denoted $f_{\rm N}$. From
Fig. \ref{ballSEAL.pdf} we get
$$F_{\rm N} = 2 \pi r_0 f_{\rm N} {\rm cos}\theta$$
$$R {\rm sin}\theta = r_0$$
so that
$$f_{\rm N}={F_{\rm N} \over 2 \pi R  {\rm cos}\theta {\rm sin}\theta}$$
Using $F_{\rm N} = \pi r_0^2 p_{\rm fluid}$ we get
$$f_{\rm N} = {1\over 2} R p_{\rm fluid}  {{\rm sin}\theta \over  {\rm cos}\theta}\eqno(1)$$

\subsection{Hertzian pressure profile}

If we assume that the contact is Hertz-like we get the pressure distribution\cite{Johnson, C4}
$$p=p_0 \left (1-\left ({x\over a}\right )^2 \right )^{1/2}\eqno(2)$$
where
$$p_0 = \left ({E^* f_{\rm N} \over \pi R}\right )^{1/2}\eqno(3)$$
$$a=\left (R\delta \right )^{1/2}\eqno(4)$$
$$f_{\rm N}={\pi \over 4} E^* \delta\eqno(5)$$
where $E^*$ is the effective Young's modulus defined by
$${1\over E^*} = {1-\nu_1^2\over E_1}+ {1-\nu_2^2\over E_2}$$
where $E_1$ and $\nu_1$ are the Young's modulus and Poisson ratio of the steel seat, and $E_2$ and $\nu_2$
the same quantities for the steel ball.

Using (1) and (3) gives
$$p_0 = \left ({E^* p_{\rm fluid}  \over 2 \pi} {{\rm sin}\theta \over  {\rm cos}\theta }\right )^{1/2}\eqno(6)$$
and from (1), (4) and (5)
$$a= R \left ({2\over \pi} {p_{\rm fluid}  \over E^*}{{\rm sin}\theta \over {\rm cos} \theta}\right )^{1/2}\eqno(7)$$
Assuming that both the steel ball and the steel cone (seat) have $E=210 \ {\rm GPa}$, $\nu = 0.3$ we get
$$E^* = {1\over 2} {E\over 1-\nu^2} \approx 115 \ {\rm GPa}$$
As an example, if $p_{\rm fluid} = 20 \ {\rm bar}$, $\theta = 45^\circ$, and $R= 2 \ {\rm cm}$ we get 
$F_{\rm N} \approx 1260 \ {\rm N}$, 
and $p_0 \approx 181 \ {\rm MPa}$ and the half-width $a\approx 0.07 \ {\rm mm}$.

\subsection{Gaussian pressure profile}

When a cylinder with a smooth surface is squeezed against a flat smooth substrate,
a rectangular contact region of width $2a$ is formed with a contact pressure given by the Hertz theory.
However, if the substrate has surface roughness the nominal contact region will be larger than
predicted by the Hertzian theory and the pressure distribution will change from parabolic-like for the
case of smooth surfaces to Gaussian-like if the surface roughness is large enough. This can be easily
shown using the Persson contact mechanics theory. Due to the surface roughness, if the contact pressure
$p$ is not too high the interfacial separation $u$ is related to the contact pressure as \cite{Yang1}
$$p=p_{\rm c} e^{-u/ u_0}\eqno(8)$$
where $u_0 = \gamma h_{\rm rms}$, where $\gamma \approx 0.4$ and where $h_{\rm rms}$
is the root-mean-square (rms) roughness amplitude. 

For a cylinder with radius $R$ which is squeezed against the flat surface we expect (see Fig.~\ref{TwoLimits.pdf})
$$u \approx u_1+{x^2 \over 2R}$$
so that
$$p=p_0 e^{-x^2/2s^2}. \eqno(9)$$
where $s^2 = \gamma R h_{\rm rms}$. 
Using (9) we get
$$\int_{-\infty}^\infty dx \ p_0 e^{-x^2/2s^2} = p_0 s ( 2\pi )^{1/2} = f_{\rm N}$$
or
$$p_0 = {f_{\rm N} \over s (2\pi )^{1/2}}\eqno(10)$$

We note that (9) holds only as long as the pressure $p$ is so small that the asymptotic relation
(8) is valid, but not too small, because then finite size effects become important.
In addition, while deriving (9) we have neglected bulk deformations. This is a valid approximation
only if $s\gg a$ or $h_{\rm rms} \gg \delta$.

Using $h_{\rm rms} = 1.9 \ {\rm \mu m}$ and $R=2 \ {\rm cm}$ gives the standard deviation $s=0.123 \ {\rm mm}$. 
A numerical study using the full $p=p(u)$ relation instead of the asymptotic
relation (8), and including bulk deformations, gives a nearly Gaussian stress distribution 
with the standard deviation $s\approx 0.139 \ {\rm mm}$ and the 
full width at half maximum (FWHM) $\approx 0.33 \ {\rm mm}$ 
(see Fig.~\ref{1x.2pcont.used.region.pdf}). These results are close to the
prediction $s= (\gamma R h_{\rm rms})^{1/2} \approx 0.123 \ {\rm mm}$, and the FWHM 
expected for a Gaussian function, which is ${\rm FWHM} = 2 (2 {\rm ln} 2)^{1/2} s \approx 2.355 s \approx 0.29  \ {\rm mm}$.

Using (1) and (10) we get
$$p_0 = {R\over 2 s (2\pi )^{1/2}} {{\rm cos}\theta \over {\rm sin}\theta} p_{\rm fluid}\eqno(11)$$
For $p_{\rm fluid} = 20 \ {\rm bar}$ and  $s= 0.123 \ {\rm mm}$
we get the maximum nominal contact pressure $p_0 \approx 57.4 \ {\rm MPa}$.


Note that the asperities act like a compliant layer on the surface of the body, so that contact is extended 
over a larger area than it would be if the surfaces were smooth and, in consequence, 
the contact pressure for a given load will be reduced.
In reality the contact area has a ragged edge which makes its measurement subject to uncertainty. 
However, the rather arbitrary definition of the contact width is not a problem when calculating physical quantities like the
leakage rate, which can be written as an integral involving the nominal pressure distribution.

\subsection{Role of plastic deformation}

The derivation of the nominal contact pressure profile (9) (and (2)) assumes elastic deformations. 
The stress-strain curve for the steel 1.4122 used for the seat shows that the stress at the onset of plastic flow (in elongation) 
is at about $500 \ {\rm MPa}$, which is much higher than the maximum stress $p_0$ (and maximum shear stress) 
at the surface and also below the surface
in the ball-seat contact region. Thus for smooth surfaces we expect no macroscopic plastic deformations, and can treat the contact as
elastic when calculating the nominal contact pressure distribution.
However, the stress in the asperity contact regions is much higher than the nominal contact pressure. 
Thus, using the power spectrum of the sandblasted seat, and including just the roughness components with 
wavelength $\lambda > 2 \ {\rm \mu m}$, and assuming elastic contact, gives for the nominal contact pressure $p_0 \approx 57 \ {\rm MPa}$
the relative contact area \cite{Per1} $A/A_0 \approx (2/h') (p_0/E^*) \approx 0.003$, where $h'$ is the rms-slope. 
Since the average pressure $p$ in the asperity contact regions
must satisfy $p A = p_0 A_0$ or $p=(A_0/A) p_0\approx h'E^*/2$ we get $p \approx 23 \ {\rm GPa}$. 
According to Tabor \cite{Tabor} the penetration hardness is $\sigma_{\rm P} \approx 3 \sigma_{\rm Y}$, where $\sigma_{\rm Y}$ is the (physical) 
yield stress in tension at about $15\%$ strain, which is about $1\ {\rm GPa}$ for the steel 1.4122.
Thus $\sigma_{\rm P} \approx 3 \ {\rm GPa}$ (we use $\sigma_{\rm P} \approx 3.5 \ {\rm GPa}$ in the calculations presented below).
We conclude that the asperities on the seat surface will deform plastically as also observed in optical pictures
of the seat surface after removing the steel ball.  

However, the derivation of (9) may still be approximately valid if the asperities deform elastically on the length scale which
determines the contact stiffness for the (nominal) contact pressures relevant for the calculation of (9). The contact stiffness
(or the $p(u)$ relation) for small pressures is determined by the most long-wavelength roughness components
which deform mainly elastically (see Sec. 4). Nevertheless, 
a more detail study is necessary in order to determine the exact influence of plastic flow at the asperity level on the nominal
contact pressure profile.

\subsection{Leak-rate theory}

In calculating the fluid (here water) leak-rate we have used the effective medium approach combined
with the Persson contact mechanics theory for the probability distribution of surface separations.
The most important region for the sealing is a narrow strip at the center of the nominal contact pressure profile,
where the contact pressure is highest (and the surface separation smallest), but the study presented below
takes into account the full pressure profile $p(x)$.

The basic contact mechanics picture which can be used to estimate the leak-rate
of seals is as follows: Consider first a seal where the nominal contact area 
is a square. The seal separates a high-pressure fluid
on one side from a low pressure fluid on the other side, with the pressure drop $\Delta P$.
We consider the interface between the solids at increasing
magnification $\zeta$. At low magnification we observe no surface roughness and it appears
as if the contact is complete. Thus studying the interface only at this low 
magnification we would be tempted to conclude that the leak-rate
vanishes. However, as we increase the magnification $\zeta$ 
we observe surface roughness and non-contact regions, so that
the contact area $A(\zeta)$ is smaller than the nominal contact area $A_0 = A(1)$. As we increase
the magnification further, we observe shorter wavelength roughness, and $A(\zeta)$ decreases
further. For randomly rough surfaces, as a function of increasing magnification, when
$A(\zeta)/A_0 \approx 0.42$ the non-contact area percolate \cite{Mueser}, and the first open channel is observed,
which allow fluid to flow from the high pressure side to the low pressure side.
The percolating channel has a most narrow constriction over which most of the pressure drop
$\Delta P$ occurs. In the simplest picture one assumes that the whole
pressure drop $\Delta P$ occurs over this {\it critical constriction}, and if it is
approximated by a rectangular pore of height $u_{\rm c}$ much smaller than its width $w$
(as predicted by contact mechanics theory), the leak rate can be approximated by\cite{Yang,Mus1,Mus2} 
$$ \dot Q = {u_{\rm c}^3 \over 12 \eta}\Delta P\eqno(12)$$
where $\eta$ is the fluid viscosity. The height $u_{\rm c}$ of the critical constriction can
be obtained using the Persson contact mechanics theory 
(see Ref. \cite{BP,Yang,Boris,Yang1,LP1,Alm,Per1}). The result (12)
is for a seal with a square nominal contact area. For a rectangular contact area with the length
$L_x$ in the fluid flow direction and $L_y$ in the orthogonal direction, there will be an
additional factor of $L_y/L_x$ in (12). In a typical case the seal has a circular (radius $r_0$) 
cross section (like for rubber O-rings), and in this case $L_y=2 \pi r_0$ and typically $L_y/L_x >> 1$ in which case the
leak-rate will be much larger than given by the {\it square-leak-rate} formula (12). However, this
geometrical correction factor is trivially accounted for. In deriving (12) it is assumed that the fluid
pressure is negligible compared to the nominal contact pressure. If this is not the case one must include the
deformation of the elastic solids by the fluid pressure distribution. This topic was studied in Ref.~\cite{liftoff}.

A more general and accurate 
derivation of the leak-rate is based on the concept of fluid flow conductivity
$\sigma_{\rm eff}$. The fluid flow
current
$$J_x = - \sigma_{\rm eff} {d p_{\rm fluid} \over d x}$$
Since the leak-rate $\dot Q = L_y J_x$ we get
$${d p_{\rm fluid} \over d x} = -{\dot Q \over L_y} {1\over \sigma_{\rm eff}}\eqno(13)$$
Note that $\sigma_{\rm eff}$ depends on the contact pressure $p_{\rm con}(x)$ and 
hence on $x$. In the present case $p_{\rm fluid} \ll p_{\rm cont}$ and in this case $p_{\rm cont} \approx p$,
where $p(x)$ is the external applied squeezing pressure (or nominal pressure) given by (9).
Integrating (13) over $x$  gives
$$\Delta P = {\dot Q \over L_y} \int_{-\infty}^{\infty} dx \ {1\over \sigma_{\rm eff}(p(x))}$$
where we have used that $\dot Q$ is independent of $x$ as a result of fluid volume conservation.
For the Hertz contact pressure profile, where $p(x) = 0$ for $x>a$ and $x<a$, 
we get with $y=x/a$:
$$\Delta P = \dot Q {2a \over L_y} \int_0^1 dy \ {1\over \sigma_{\rm eff}(p (y))}$$
where
$$p_{\rm c} (y)=p_0 \left (1-y^2\right )^{1/2}$$

For the Gaussian pressure profile (9) using $y=x/s$ we get
$$\Delta P = \dot Q {2s \over L_y} \int_0^\infty dy \ {1\over \sigma_{\rm eff}(p (y))}\eqno(14)$$
where
$$p (y)=p_0 e^{-y^2/2}.$$
From (14) we get the fluid leak-rate (volume per unit time)
$$\dot Q = {L_y \over 2s} {\Delta P \over \int_0^\infty dy \ \sigma^{-1}_{\rm eff}(p (y))}\eqno(15)$$


In the calculations presented below we have not used the critical junction theory but 
Eq.~(15) with the flow conductivity $\sigma_{\rm eff}$ calculated using
the Bruggeman effective medium theory (``corrected'' so the contact area percolate for $A/A_0 = 0.42$; see Ref.~\cite{Mueser}),
and the Persson contact mechanics theory (see Ref. \cite{Boris,LP1} for the details). 
This theory takes into account all the fluid flow channels and not
just the first percolating channel observed with increasing magnification as in the critical junction theory. 
The dependency of the leak-rate on the fluid viscosity $\eta$ and
the fluid pressure difference $\Delta P$ given by (11) is the same in the more accurate approach. 
Similar, the leak-rate is proportional to $L_y/L_x$ (where $L_x=2s$ in the present case) 
in this more accurate approach. In fact, the Bruggeman effective
medium theory and the critical junction theory gives in most cases very similar numerical results.
Comparison of the prediction of the Bruggeman effective medium theory for the leak-rate with exact numerical studies
has shown the effective medium theory to be remarkably accurate in the present context \cite{Mueser}.

\subsection{Accounting for plastic deformations}

In the present study we are interested in metallic seals and in this case plastic deformation of the
solids is very important. Plastic flow is a complex topic but two very simple approaches have been proposed to
take into account plastic deformations in the context of metallic seals. One approach,
which is simple to implement when using actual realizations of the rough surfaces
(as done in most numerically treatments, e.g., using the boundary
element method \cite{Alm}), is to move surface grid points 
vertically in such a way that the stress in the plastically deformed region 
is equal to (or below) the penetration hardness.

Another approach, which is more convenient in analytic contact mechanics theories, is based
on smoothing the surface in wavevector space (see \cite{Yang,Boris,LP1}). 
Thus, if two solids are squeezed together with the pressure
$p_0$ they will deform elastically and, at short enough
length scale, plastically. If the contact is now removed the
surfaces will be locally plastically deformed. Assume now
that the surfaces are moved into contact again at exactly
the same position as the original contact, and with the
same squeezing pressure $p_0$ applied. In this case the solids
will deform purely elastically and the Persson contact mechanics theory 
can be (approximately) applied assuming that
the surface roughness power spectrum $C_{\rm pl}(q)$ of the (plastically) 
deformed surface is known. 

An expression for $C_{\rm pl}(q)$ can be obtained as follows.
Let us consider the contact between two elastoplastic bodies with nominal flat surfaces,
but with surface roughness extending over many decades in length scale, as is almost always the case.
Assume that the applied (nominal) contact pressure $p_0$ is smaller than the penetration hardness $\sigma_{\rm P}$
of the solids. When we study the contact between the solids at low magnification we do not observe any surface roughness,
and since $p_0 < \sigma_{\rm P}$ the solids deform purely elastically at this length scale. As we increase the
magnification we observe surface roughness and the (elastic) contact area decreases. At some magnification
the pressure $p=p_0 A_0/A(\zeta)$ may reach the penetration hardness and at this point all the the contact regions are
plastically deformed. In general, depending on the magnification $\zeta$, some fraction of the contact area involves elastic
deformations, while the other fraction has undergone plastic deformation. Thus we can write the contact area
$A(\zeta ) = A_{\rm el}(\zeta)+A_{\rm pl}(\zeta)$. The Persson contact mechanics theory predicts both
$A_{\rm el}(\zeta)$ and $A_{\rm pl}(\zeta)$ (see Ref. \cite{BP}).

In ref.~\cite{roughness} we have shown
that $C_{\rm pl}(q)$ can be obtained approximately using (with $\zeta = q/q_0$, where $q_0$ is the smallest wavenumber) 
$$C_{\rm pl} (q)= \left [1-\left ({A_{\rm pl} (\zeta) \over A^0_{\rm pl}}\right )^6\right ]C(q),\eqno(16)$$ 
where $A^0_{\rm pl}=F_{\rm N}/\sigma_{\rm P}$ is the contact area assuming that all contact regions 
have yielded plastically so the pressure
in all contact regions equal the penetration hardness $\sigma_{\rm P}$.
The basic picture behind this definition is that 
surface roughness at short length scales gets
smoothed out by plastic deformation, resulting in an effective cut-off 
of the power spectrum for large wave vectors (corresponding to short distances).
Assuming elastic contact and using the power spectrum (16) result in virtually the same (numerical) contact area $A(\zeta)$, as a function of
magnification $\zeta$, as predicted for the original surface using the elastoplastic contact mechanics theory,
where $A(\zeta ) = A_{\rm el}(\zeta)+A_{\rm pl}(\zeta)$.

The smoothing of the surface profile at short length scale allows the surfaces
to approach each other and will reduce the height $u_{\rm c}$ of the critical constriction.
By using the plastically deformed surface roughness power spectrum (16) this effect is taken into
account in a simple approximate way.

It is not clear which of the two ways to include plastic flow described above is most accurate. 
In the first approach the solids deform plastically only in the region where they make contact,
but this procedure does not conserve the volume of the solids. The second approach does conserve the volume
but smooth the surfaces everywhere, i.e., even in the non-contact region. However, this does not influence the area
of real contact, and it has a relative small influence on the surface separations in the big fluid flow channels, which mainly
determine the fluid leakage rate (see Ref. \cite{Per1} for a discussion of this).

\begin{figure}
\includegraphics[width=0.95\columnwidth]{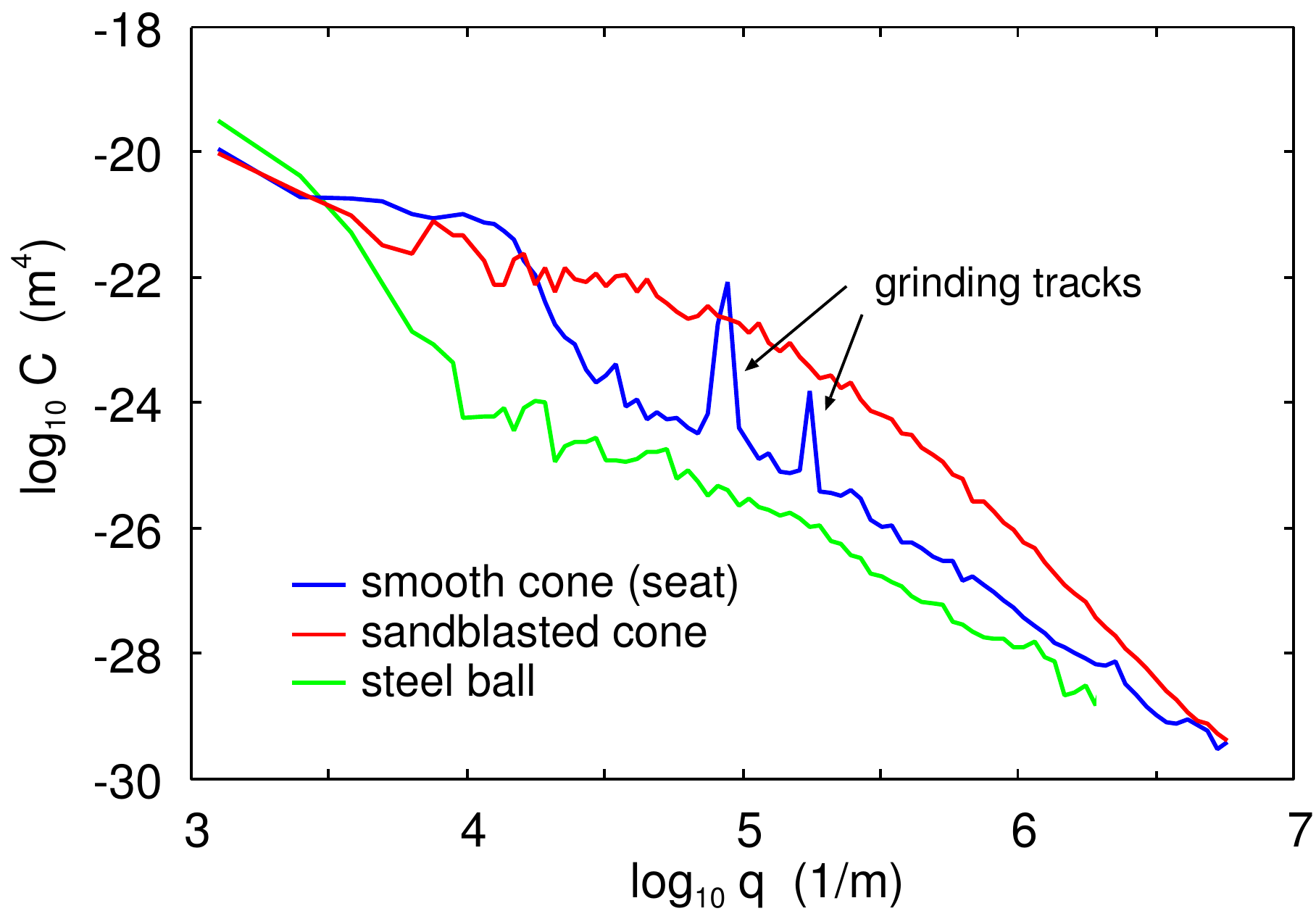}
\caption{\label{1logq.2logC.red59SandblastedSeal.blue9SmoothSeal.green13Ball.pdf}
The (measured) 2D surface roughness power spectrum of the smooth steel cone (seat) surface
(blue line), of the sandblasted cone surface (red line) and of the steel ball (green line).
The rms roughness are $1.2 \ {\rm \mu m}$, $1.9 \ {\rm \mu m}$ and $0.8 \ {\rm \mu m}$,
respectively. The rough cone has the rms slope $0.4$ (isotropic), 
and the smooth cone $0.11$ in the circumferential direction, and $0.18$ in the radial direction.
}
\end{figure}

\begin{figure}
\includegraphics[width=0.95\columnwidth]{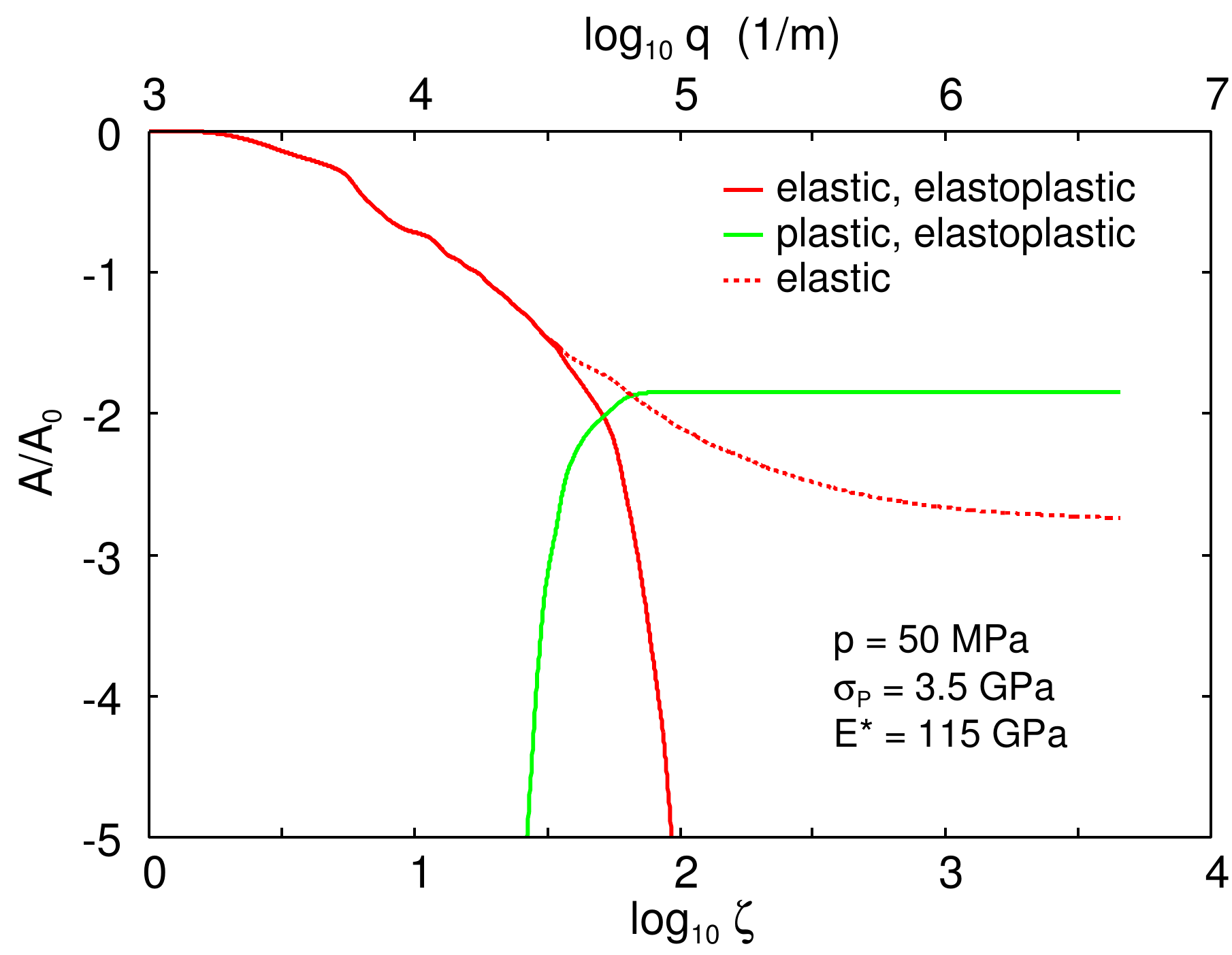}
\caption{\label{1logz.2logArea.all.pdf}
The relative elastic $A_{\rm el}/A_0$ (red line) and relative plastic $A_{\rm pl}/A_0$ (green line)
contact area as a function of the magnification $\zeta$. The red dotted line is the relative contact
area without plasticity. In the elastoplastic calculation we use the penetration hardness $\sigma_{\rm P} = 3.5 \ {\rm GPa}$.
For the effective Young's modulus $E^* = 115 \ {\rm GPa}$ and the nominal contact pressure $p_0=50 \ {\rm MPa}$.
}
\end{figure}

\begin{figure}
\includegraphics[width=0.95\columnwidth]{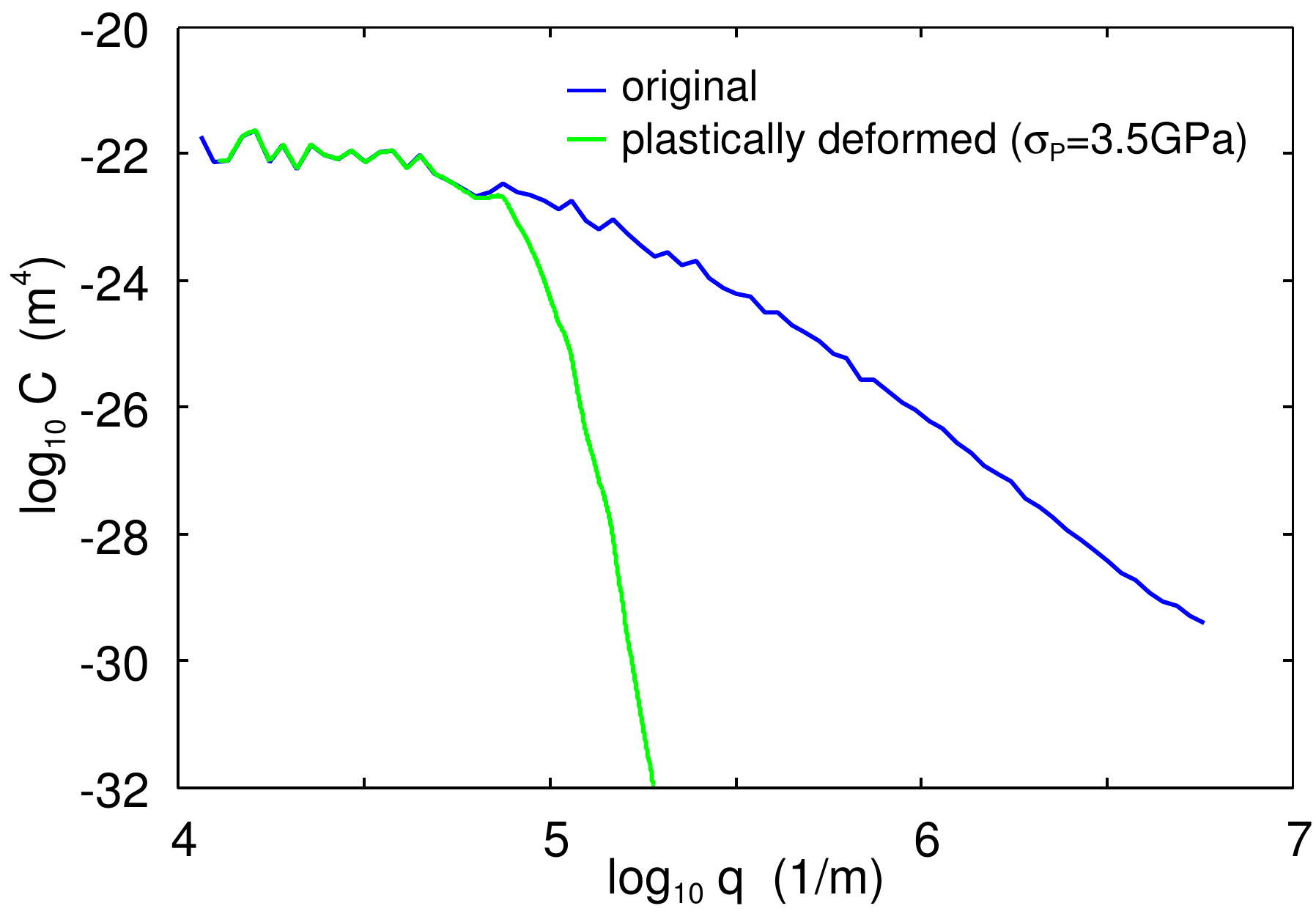}
\caption{\label{logq.2logC.used.59.pdf}
The (measured) 2D surface roughness power spectrum of the sandblasted steel cone (seat) surface
(blue line), and the (calculated) power spectrum of the plastically deformed cone
surface (green solid line).
}
\end{figure}

\begin{figure}
\includegraphics[width=0.95\columnwidth]{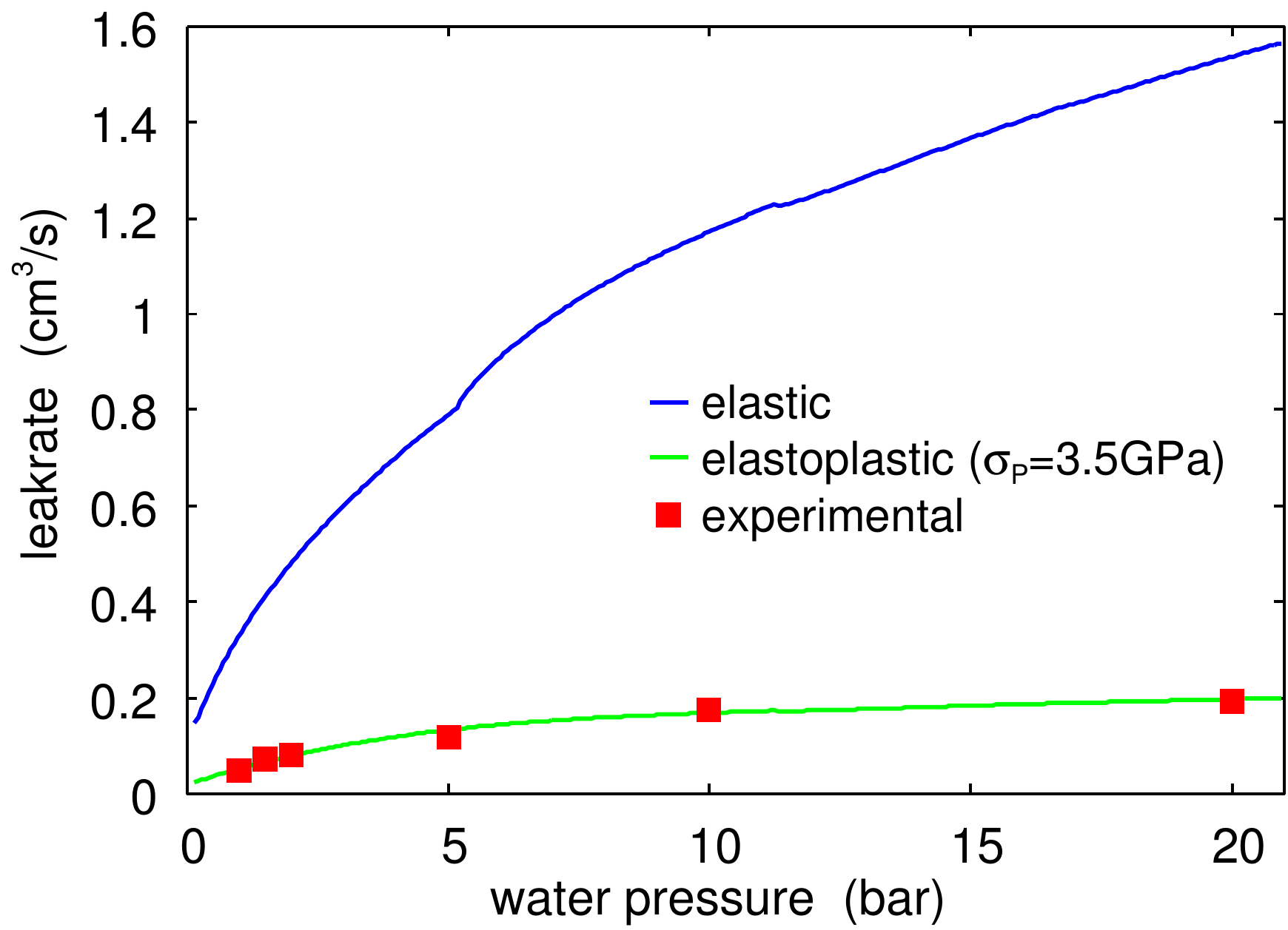}
\caption{\label{x1pwater.2lekagage.metallic.seal.rough.seal.59.pdf}
The measured (red squares) and the calculated water leak-rate as a function of the water
pressure difference. The green line is the result including plastic deformation, and the blue line
assuming only elastic deformation.
The steel ball is squeezed against the sandblasted conical surface (seat) only by the water pressure
so increasing the water pressure also increases the normal force squeezing the ball against the seat's surface.
}
\end{figure}

\begin{figure}
\includegraphics[width=0.95\columnwidth]{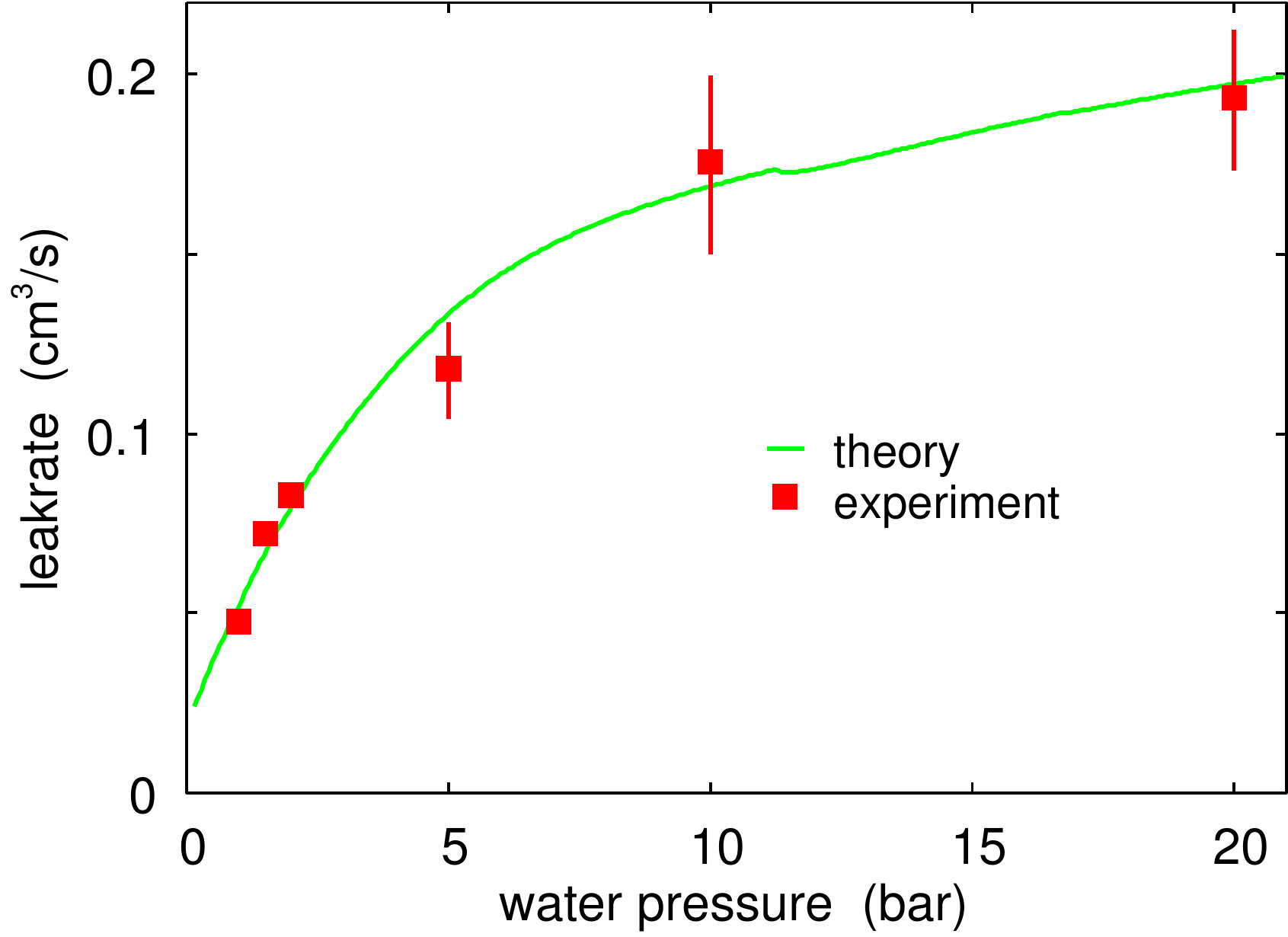}
\caption{\label{1pwater.2dorQ.plastic.GAUSS.3.5GPa.0.5mm.pdf}
The measured (red squares) and the calculated (green line) water leak-rate as a function of the water
pressure difference. The theory curve includes the plastic deformation.
}
\end{figure}

\begin{figure}
\includegraphics[width=0.95\columnwidth]{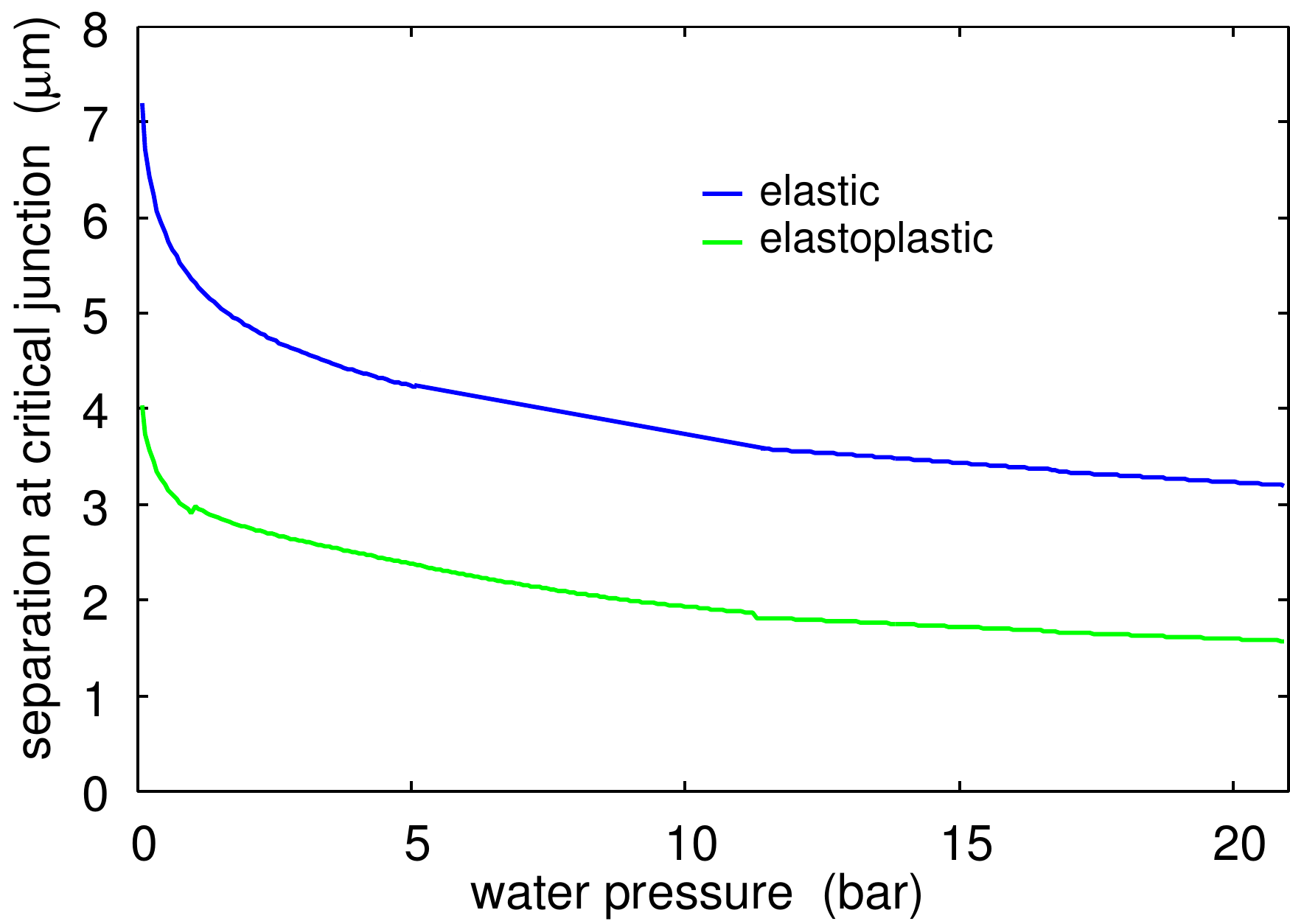}
\caption{\label{1waterp.2separation.cj.pdf}
The calculated surface separation at the critical junction as a function of the water pressure.
The green line is the result including plastic deformation, and the blue line
assuming only elastic deformation.
}
\end{figure}

\begin{figure}
\includegraphics[width=0.95\columnwidth]{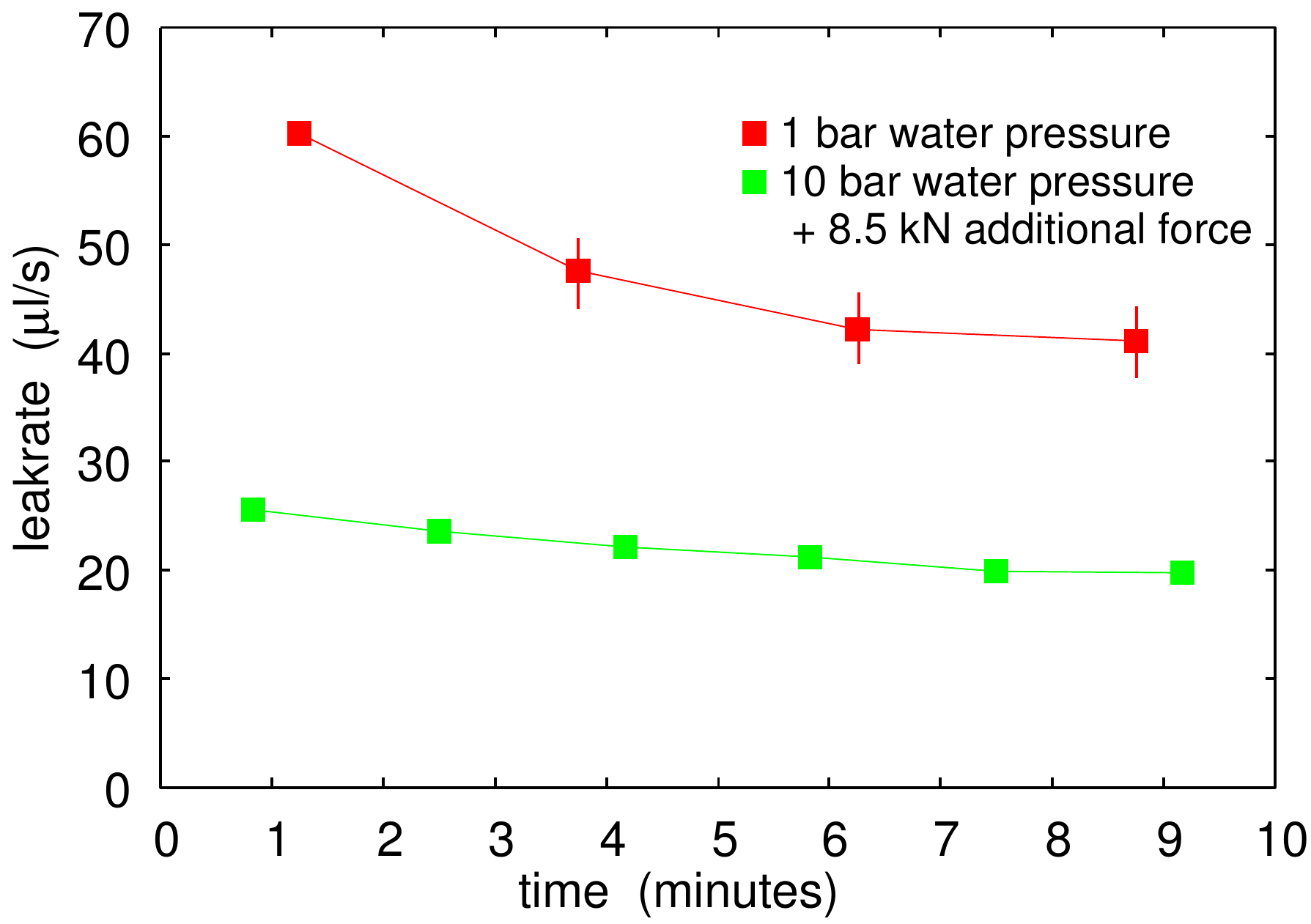}
\caption{\label{times.leakrate.4.pdf}
The time dependency of the measured leakrate for the sandblasted seat at the water pressure $p_{\rm water} = 1 \ {\rm bar}$
(red squares) and $p_{\rm water} = 10 \ {\rm bar}$ (green squares). 
For the $p_{\rm water} = 10 \ {\rm bar}$ case an additional normal force of $8.5 \ {\rm kN}$ was squeezing the ball against the seat.
The decrease in the leakage rate with increasing time is likely
due to clogging of flow channels by contamination particles.
}
\end{figure}

\section{Experimental results and analysis}

Fig.~\ref{1logq.2logC.red59SandblastedSeal.blue9SmoothSeal.green13Ball.pdf}
shows the (measured) 2D surface roughness power spectrum of the smooth steel cone's (seat) surface
(blue line), of the sandblasted seat's surface (red line) and of the steel ball (green line).
The rms roughness are $1.2 \ {\rm \mu m}$, $1.9 \ {\rm \mu m}$ and $0.8 \ {\rm \mu m}$,
respectively. The rough cone has the rms slope $0.4$ (isotropic), 
and the smooth seat $0.11$ in the circumferential direction, and $0.18$ in the radial direction.
The smooth seat has anisotropic roughness, resulting from the grinding process which introduced
wear tracks in the circumferential direction. These wear tracks are easily observed with the naked eyes,
and shows up in the power spectrum in the radial direction as the two sharp
peaks around $q\approx 10^5 \ {\rm m}^{-1}$. Since the theory we use to analyze the experimental data assumes
randomly rough surfaces (but it can be generalized to include periodic roughness \cite{Scaraggi}),
in this paper we consider only the seat with the sandblasted surface (red line), which is randomly rough to a good approximation.

Fig.~\ref{1logz.2logArea.all.pdf}
shows the (calculated) relative elastic $A_{\rm el}/A_0$ (red line) and relative plastic $A_{\rm pl}/A_0$ (green line)
contact area as a function of the magnification $\zeta$ (lower log-scale) and the wavenumber $q$ (upper log-scale). 
The red dotted line is the relative contact
area without plasticity. In the elastoplastic calculation we use the penetration hardness $\sigma_{\rm P} = 3.5 \ {\rm GPa}$,
and the effective Young's modulus $E^* = 115 \ {\rm GPa}$ and the nominal contact pressure $p_0=50 \ {\rm MPa}$.
Note that the long wavelength roughness is elastically deformed, but already at the
magnification $\zeta \approx 100$ all the contact area is observed to be plastically deformed. The magnification $\zeta \approx 100$
corresponds to the wavenumber $q=\zeta q_0 \approx 10^5 \ {\rm m}^{-1}$ or the wavelength $\lambda = 2\pi /q \approx 60 \ {\rm \mu m}$.
Thus all the contact regions observed with, e.g., an optical microscope (with the optimal resolution determined by the
wavelength of light, $\lambda \approx 1  \ {\rm \mu m}$), are plastically deformed. 

Fig.~\ref{logq.2logC.used.59.pdf}
shows the (measured) 2D surface roughness power spectrum of the sandblasted steel seat's surface
(red line, from Fig. \ref{1logq.2logC.red59SandblastedSeal.blue9SmoothSeal.green13Ball.pdf}), 
and the (calculated) power spectrum of the plastically deformed seat's (green solid line) as obtained using (16).  

Fig.~\ref{x1pwater.2lekagage.metallic.seal.rough.seal.59.pdf}
shows the measured (red squares) and the calculated water leak-rate as a function of the water
pressure difference. The green line is the result including plastic deformation, and the blue line
assuming only elastic deformation.
The steel ball is squeezed against the sandblasted seat only by the water pressure
so increasing the water pressure also increases the normal force squeezing the ball against the seat's surface.
This will reduce the interfacial separation at the critical junctions, and hence the leakrate, and this explains
why the leakrate does not increase proportional to the fluid pressure difference $\Delta P$ as otherwise expected.

Note that including the plastic deformation results in a drastic reduction, by roughly a factor of 8, 
in the (calculated) fluid leak-rate. The theory prediction with the plastic deformation included is in very good agreement with
experiments. This is shown in greater detail in Fig.~\ref{1pwater.2dorQ.plastic.GAUSS.3.5GPa.0.5mm.pdf}.
We note that all the parameters which enter in the theory, like surface roughness power spectrum or elastoplastic 
modulus, have been measured directly so there is no fitting parameter in the theory calculation.

The leak-rate depends on the separation at the critical constriction as $u_{\rm c}^3$. Thus the reduction in the leak-rate by a 
factor of $\approx 8$ when including the plastic deformation imply that the separation $u_{\rm c}$ decreases with a factor of $\approx 2$.
This is in good agreement with the result shown in Fig. \ref{1waterp.2separation.cj.pdf} which shows
the calculated surface separation at the critical junction as a function of the water pressure.
The green line is the result including plastic deformation, and the blue line
assuming only elastic deformation. 

Fig.~\ref{1waterp.2separation.cj.pdf} shows that if there would be contamination particles in the fluid of
micrometer size they could clog the flow channels and hence reduce the fluid leakage rate with increasing time. In spite of the
fact we used purified water we did observe a dependency of the leak-rate on time. This may be due to dust particles in the water,
which could not be completely avoided as the experiments was performed in the normal atmosphere. Other possible reasons for the contamination are internal leakage from the air side or the oil side or corrosion of the materials.
To illustrate this effect, in Fig.~\ref{times.leakrate.4.pdf} 
we show the measured time dependency of the measured leak-rate for the sandblasted seat at the water pressure $p_{\rm water} = 1 \ {\rm bar}$
(red squares) and $p_{\rm water} = 10 \ {\rm bar}$ (green squares). 
The steel ball is squeezed against the smooth seat by the water pressure,
but for the $p_{\rm water} = 10 \ {\rm bar}$ case an additional normal force of $8.5 \ {\rm kN}$ was applied via the piston in Fig.~\ref{ballseatAachen.pdf}. The decrease in the leakage rate with increasing time may be due to 
clogging of flow channels by contamination particles.

\section{Summary and conclusion}

We have investigated the role of plastic deformation in the leak-rate of metallic seals.
We found that plastic deformation increases the area of real contact and reduces the
interfacial separation at the critical constriction, which reduces the leak rate by roughly a factor of 8.
Our experimental results show a nonlinear dependency of leak-rate with fluid pressure difference, due to
a dependency of the applied axial force on the fluid pressure. The theoretical results,
based on the Persson's contact mechanics theory in combination with the 
Bruggeman effective medium theory are in good agreement with measured data for the leak-rate as a function of the fluid pressure. The measured leak-rate decreases with increasing time, which we interpret as resulting from
clogging of critical constrictions by impurities present in the water.

\section*{Acknowledgments}
This work was funded by the German Research Foundation (DFG) in the scope of the 
Project "Modellbildung metallischer Dichtsitze" (MU1225/42-1). The authors would like to thank DFG for its support.


\begin{thebibliography}{99}

\bibitem{add1}
What determines seal leakage? - Fluid Sealing Association        
(May 2008) see: 

http://www.fluidsealing.com/sealingsense/may08.pdf

\bibitem{armand}
G. Armand,  J. Lapujoulade and J. Paigne: A theoretical and experimental 
relationship between the leakage of gases through the interface of two metals in contact and their superficial micro-geometry, Vacuum 14, 53 (1964).

\bibitem{thesis}
D. Nurhadiyanto: Influence of surface roughness on leakage of corrugated 
metal gasket, Dissertation, Yamaguchi University, Japan, September 2014;
See also: S. Haruyama, D. Nurhadiyanto and M.A. Choiron and K. Kaminishi,
Influence of surface roughness on leakage of new metal gasket,
International Journal of Pressure Vessels and Piping {\bf 111-112}, 146 (2013).

\bibitem{thesis1}
F. Perez-Rafols:
Modelling and numerical analysis of leakage through metal-to-metal seals,
Licentiate thesis,
Lulea University of Technology
Department of Engineering Science and Mathematics,
Division of Machine Elements (2016).
see:

http://pure.ltu.se/portal/files/105667647/

Francesci$\_$P$\_$rez$\_$R$\_$fols.pdf

\bibitem{metals}
F. Perez-Rafols, R. Larsson and A. Almqvist,
Modelling of leakage on metal-to-metal seals,
Tribology International 94, 421 (2016).

\bibitem{Lorenz1}
B. Lorenz, B.N.J. Persson,
{\it Leak rate of seals: Effective-medium theory and comparison with experiment},
The European Physical Journal E {\bf 31}, 159 (2010)

\bibitem{Lorenz2}
B. Lorenz, B.N.J. Persson,
{\it Leak rate of seals: Comparison of theory with experiment},
Europhysics Letters {\bf 86}, 44006 (2009).


\bibitem{Ref7}
Martin H M\"user, Wolf B Dapp, Romain Bugnicourt, Philippe Sainsot, Nicolas Lesaffre, Ton A Lubrecht, Bo NJ Persson, Kathryn Harris, Alexander Bennett, Kyle Schulze, Sean Rohde, Peter Ifju, W Gregory Sawyer,
Thomas Angelini, Hossein Ashtari Esfahani, Mahmoud Kadkhodaei, Saleh Akbarzadeh, Jiunn-Jong Wu, Georg Vorlaufer, Andras Vernes, Soheil Solhjoo,
Antonis I Vakis, Robert L Jackson, Yang Xu, Jeffrey Streator, Amir Rostami, Daniele Dini,
Simon Medina, Giuseppe Carbone, Francesco Bottiglione, Luciano Afferrante, Joseph Monti, Lars Pastewka, Mark O Robbins, James A Greenwood,
{\it Meeting the contact-mechanics challenge},
Tribology Letters {\bf 65}, 118 (2017).

\bibitem{BP}
BNJ Persson
{Theory of rubber friction and contact mechanics},
The Journal of Chemical Physics {\bf 115}, 3840 (2001)


\bibitem{Alm}
A. Almqvist, C. Campan, N. Prodanov and B.N.J. Persson,
{\it Interfacial separation between elastic solids with randomly rough
	surfaces: Comparison between theory and numerical techniques},
Journal of the Mechanics and Physics of Solids {\bf 59}, 2355 (2011).

\bibitem{Per1}
B.N.J. Persson,
{\it Leakage of metallic seals: role of plastic deformations},
Tribology Letters {\bf 63}, 42 (2016).


\bibitem{AA4}
F. Per\'{e}z R\`{a}fols, R.  Larsson, S. Lundstr\"om, P.  Wall, A. Almqvist,
{\it A stochastic two-scale model for pressure-driven flow between rough surfaces}, 
Proceedings of the Royal Society A: Mathematical, Physical and Engineering Sciences, 472(2190), 20160069 (2016).


\bibitem{Yang}
B.N.J. Persson and C. Yang:
Theory of the leak-rate of seals,
Journal of Physics, condensed matter {\bf 20}, 315011 (2008).

\bibitem{Review}
B.N.J. Persson, O. Albohr, U. Tartaglino, A.I. Volokitin, E. Tosatti,
On the nature of surface roughness with application to contact mechanics, sealing, rubber friction and adhesion
Journal of Physics: Condensed Matter {\bf 17}, R1 (2004).

\bibitem{Boris}
B. Lorenz and B.N.J. Persson: 
Leak rate of seals: Effective-medium theory and comparison with experiment,
European Physics Journal E {\bf 31}, 159 (2010). 

\bibitem{LP1}
B. Lorenz, B.N.J Persson
Leak rate of seals: Comparison of theory with experiment,
EPL {\bf 86}, 44006 (2009).

\bibitem{liftoff}
B. Lorenz and B.N.J. Persson
Time-dependent fluid squeeze-out between solids with rough surfaces
The European Physical Journal E {\bf 32}, 281 (2010).


\bibitem{Mueser}
W.B. Dapp, A. Lucke, B.N.J. Persson and M.H. M\"user:  
Self-Affine Elastic Contacts: Percolation and Leakage,
Phys. Rev.Lett. {\bf  108}, 244301 (2012).

\bibitem{Mus1} 
W.B. Dapp and M. H. M\"user:
{\it Fluid leakage near the percolation threshold}
Sci. Rep. {\bf 6}, 19513 (2016).

\bibitem{Mus2}
W.B. Dapp and M.H. M\"user,
{\it Contact mechanics of and Reynolds flow through saddle points}
EPL {\bf 109}, 44001 (2015).

\bibitem{Carbone}
F. Bottiglione, G. Carbone, L. Mangialardi and G. Mantriota,
Leakage mechanism in flat seals,
Journal of Applied Physics {\bf 106}, 104902 (2009).

\bibitem{tobe}
F. Perez-Rafols, R. Larsson, E.J. van Riet and A. Almqvist,
On the flow through plastically deformed surfaces under loading. A frequency based approach.
(To be pulished.)

\bibitem{Pei}
L. Pei, S. Hyun, J.F. Molinari and M.O. Robbins,
Finite element modeling of elasto-plastic contact between rough surfaces
Journal of the Mechanics and Physics of Solids {\bf 53}, 2385 (2005).

\bibitem{Kadin}
Y. Kadin, Y. Kligerman and I. Etsion,
Unloading an elastic-plastic contact of rough surfaces,
Journal of the Mechanics and Physics of Solids {\bf 54}, 2652 (2006).


\bibitem{Zhao}        
B. Zhao, S. Zhang, , P. Wang and Y. Hai,
Loading-unloading normal stiffness model for power-law hardening surfaces considering actual surface topography,
Tribology International {\bf 90}, 332 (2015).

\bibitem{Av}
A. Tiwari, A. Wang, M.H. M\"user, B.N.J. Persson,
{\it Contact Mechanics for Solids with Randomly Rough Surfaces and Plasticity}
Lubricants {\bf 7}, 90 (2019).

\bibitem{Av2}
A.Tiwari, A. Almqvist. B. N. J. Persson
{\it Plastic deformation of rough metallic surfaces},
arXiv:2006.11084

\bibitem{Pas}
A.R. Hinkle, W.G. N\"ohring, R. Leute, T. Junge, L. Pastewka,
{\it The emergence of small-scale self-affine surface roughness from deformation},
Science advances {\bf 6}, 0847 (2020).

\bibitem{Nic1}
S. P.Venugopalan, M. H.  M\"user, L. Nicola,
{\it Green's function molecular dynamics meets discrete dislocation plasticity}, 
 Modelling and Simulation in Materials Science and Engineering, 25(6), 065018 (2017).

\bibitem{Nic2}
S.P. Venugopalan, N. Irani, L. Nicola,
{\it Plastic contact of self-affine surfaces: Persson's theory versus discrete dislocation plasticity},
Journal of the Mechanics and Physics of Solids {\bf 132}, 103676 (2019).

\bibitem{Nic3}
N. Irani, L. Nicola,
{\it Modelling surface roughening during plastic deformation of metal crystals under contact shear loading},
Mechanics of Materials {\bf 132}, 66 (2019).


\bibitem{Nic4}
S.P. Venugopalan, L. Nicola,
{\it Indentation of a plastically deforming metal crystal with a self-affine rigid surface: A dislocation dynamics study},
Acta Materialia 165, 709 (2019).

\bibitem{fractal}
B.N.J. Persson,
On the fractal dimension of rough surfaces,
Tribology Letters {\bf 54}, 99 (2014).


\bibitem{Nyak}
P.R. Nayak,
{\it Random Process Model of Rough Surfaces},
J. Lubrication Technol. {\bf 93}, 398 (1971).

 \bibitem{CarbLor}
G. Carbone, B. Lorenz, B.N.J. Persson, A. Wohlers,
{\it Contact mechanics and rubber friction for randomly rough surfaces with anisotropic statistical properties},
The European Physical Journal E{\bf 29}, 275 (2009).

\bibitem{Yang1}
C. Yang and B.N.J. Persson: Contact mechanics: 
contact area and interfacial separation from small contact to full contact, 
J. Phys.: Condens. Matter {\bf 20}, 215214 (2008).

\bibitem{C5}
B.N.J. Persson,
{\it Contact mechanics for randomly rough surfaces},
Surface Science Reports {\bf 61}, 201 (2006).



\bibitem{PRL}
B.N.J. Persson,
{\it Relation between Interfacial Separation and Load: A General Theory of Contact Mechanics},
Phys. Rev. Lett. {\bf 99}, 125502 (2007).

\bibitem{Johnson}
K. L. Johnson,  
{\it Contact mechanics},
 Cambridge university press (1987).
 
 \bibitem{C4}
J.R. Barber, {\it Contact Mechanics  (Solid Mechanics and Its Applications)},
 Springer (2018).

\bibitem{Tabor}
D. Tabor, 
{\it The hardness of metals},
Clarendon Press, Oxford, UK (1951).

\bibitem{roughness}
B.N.J. Persson, B. Lorenz and A.I. Volokitin,
Heat transfer between elastic solids with randomly rough surfaces
The European Physical Journal E {\bf 31}, 3 (2010); see also:
B.N.J. Persson, Surface Science Reports
{\bf 61}, 201 (2006).


\bibitem{Scaraggi}
M. Scaraggi and G. Carbone, 
{\it A Two-Scale Approach for Lubricated Soft-Contact Modeling: An Application to Lip-Seal Geometry}, 
Advances in Tribology, Article ID 412190, 1-12 (2012)


%
%
%
%

%
%
%
%
%


%
%

%
%
%
%
%




%
%






\end{thebibliography}
\end{document}